\documentclass[11pt]{article}
\usepackage[margin=2cm,nohead]{geometry}                
\usepackage{graphicx}
\usepackage{amssymb}
\usepackage{courier}
\usepackage[utf8]{inputenc}

\usepackage{amssymb}

\usepackage{amsmath}
\usepackage{stmaryrd}
\usepackage[fleqn,tbtags]{mathtools} 
\usepackage{subfig}
\usepackage{array}
\usepackage{textcomp } 
\usepackage{ifthen}
\usepackage{bbding} 
\usepackage{upquote}
\usepackage{colortbl} 

\usepackage[pdftex,
            hyperindex=true,
            colorlinks=true,
            linkcolor=blue,
            anchorcolor=magenta,
            citecolor=blue,
            urlcolor=blue,
            unicode,
            implicit=true
            ]{hyperref}

\def\ifmonospace{\ifdim\fontdimen3\font=0pt }

\def\C++{%
\ifmonospace%
    C++%
\else%
    C\kern-.1667em\raise.50ex\hbox{\tiny{\textbf{+}\kern-.1em\textbf{+}}}%
\fi%
\spacefactor1000 }

\def\CPP11{%
\ifmonospace%
    C++11%
\else%
    C\kern-.1667em\raise.50ex\hbox{\tiny{\textbf{+}\kern-.1em\textbf{+}}}\kern-.1em11%
\fi%
\spacefactor1000 } 

\definecolor{lightergray}{rgb}{0.8,0.8,0.8}

\usepackage{listings}
\usepackage[dvipsnames,usenames]{xcolor}
\usepackage{listings}
\usepackage{upquote}  

\usepackage{caption}
\DeclareCaptionFont{white}{\color{white}}
\DeclareCaptionFormat{listing}{\colorbox{gray}{\parbox{\textwidth-2cm}{#1#2#3}}}
\DeclareMathAlphabet{\mathpzc}{OT1}{pzc}{m}{it}

\newcommand{\ttfont}[1]{{\ttfamily{#1}}}
\newcommand{\code}[1]{{\footnotesize\ttfamily{#1}}}

\newcommand\keywordd[1]{{\color{DarkOrchid}\footnotesize\ttfamily\textbf{#1}}}

\graphicspath{{./figures/}}

\title{A \C++11 implementation of arbitrary-rank tensors for high-performance computing}
\author{Alejandro M. Arag\'{o}n}
\date{}                                           

\begin{document}
\maketitle

The published version of the article can be found in \href{http://dx.doi.org/10.1016/j.cpc.2014.01.005}{\emph{Computer Physics Communications}}.

\begin{abstract}

This article discusses an efficient implementation of tensors of arbitrary rank
by using some of the idioms introduced by the recently published \C++ ISO Standard (\CPP11). 
With the aims at providing a basic building block for high-performance computing,
a single \code{Array} class template is carefully crafted, from which vectors, matrices, and even higher-order tensors can be created.
An expression template facility is also built around the array class template to provide convenient mathematical syntax.
As a result, by using templates, an extra high-level layer is added to the \C++ language when dealing with algebraic objects and their operations, without compromising performance. The implementation is tested running on both CPU and GPU.
\end{abstract}


\section{Introduction} \label{sec:intro}

Fortran has been traditionally regarded as the programming language of choice for high performance computing (HPC). Nevertheless, scientists are increasingly adopting other languages motivated by the use of higher level paradigms, object orientation, data abstraction, and generic programming. One of these languages is \C++, developed in the late 1970s to enhance the C language~\cite{C99}, and standardized in 1998~\cite{C++98}.
Contrary to Fortran, the \C++ programming language does not provide built-in support for matrices.
As a result, a programmer is forced to either write custom classes that represent the abstraction of algebraic objects, or to use an external library, e.g., PETSc~\cite{PETSc}, Blitz++~\cite{Veldhuizen:1998}, or MTL~\cite{Siek:1998, Gottschling:2007}. 
In the former situation, a naive implementation could drastically deteriorate the performance of a computer program, and this unfortunate situation may even dissuade the user from using \C++ entirely.
In the latter case, the dependency of the resulting code increases, for the user has to build the external library and its dependents.
The addition of a library to handle algebraic objects also raises the important question: if an external library is needed to cope with the absence of such an elemental component in scientific computing, how many other libraries need to be added to the dependency list for a final software design?

The Basic Linear Algebra Subprograms (BLAS)~\cite{Lawson:1979, Dongarra:1988, Dongarra:1990} is the programming interface of choice when carrying out algebraic operations with vectors and matrices. A highly optimized library that implements the BLAS interface can be found in any high-performance scientific computing environment. However, the library interface is not straightforward to use, as revealed in Listing~\ref{lst:blas_dgemm} by the fourteen-parameter function that is used to multiply two matrices.
\begin{lstlisting}[caption={Level 3 BLAS cblas\_dgemm function}, label=lst:blas_dgemm]
void cblas_dgemm(const enum CBLAS_ORDER Order, const enum CBLAS_TRANSPOSE TransA,
                       const enum CBLAS_TRANSPOSE TransB, const int M, const int N, const int K,
                       const double alpha, const double *A,  const int lda, const double *B, 
                       const int ldb, const double beta, double *C, const int ldc);                 
\end{lstlisting}
The BLAS interface is consequently prone to user error, and a more friendly interface is desirable, even if an optimized implementation of the BLAS interface is still used at the backend.
One such project is the uBLAS library~\cite{ublas:2002}, that is part of the BOOST set of \C++ libraries. Yet, this library only supports vectors and matrices, and it may prove valuable in several fields of scientific computing to have even higher order tensors.

In the words of Stroustrup, the creator of \C++~\cite{Stroustrup:1994}:
\begin{quote}
\emph{It would be nice if every kind of numeric software could be written in \C++ without loss of efficiency, but unless something can be found that achieves this without compromising the \C++ type system it may be preferable to rely on Fortran, assembler, or architecture-specific extensions.}
\end{quote}
By using some of the features of the newly published \C++ standard, 
this article outlines a methodology by which numeric software can now be entirely written in \C++ in a fully type-safe environment, and without compromising performance.
Firstly, a single \code{Array} class template
is thoroughly crafted so as to create tensors of arbitrary rank. This is possible by virtue of new idioms introduced by the recently published \C++ ISO Standard~\cite{C++11}, which aims at enhancing the programming language thorough the set of \CPP11 requirements (the standard is formally known as ISO/IEC 14882:2011).
By going through the implementation of the \code{Array} class template, the reader is introduced to many of the new features in \CPP11, including variadic templates, initializer lists, lambda expressions, among others.
Secondly, the article further outlines the machinery by which it is possible to use the straightforward mathematical notation known to every computational scientist, thus providing a high-level layer when dealing with the tensors.
In other words, the end user could write \code{C=A*B} to obtain the product between two matrices \code{A} and \code{B}, even if the implementation calls the aforementioned BLAS function shown in Listing~\ref{lst:blas_dgemm}. This is possible by means of a combination of \emph{operator overloading} and \emph{expression templates}.
The end result is the powerful capability of tailoring the syntax for specific operations, so that complicated expressions are matched and forwarded to BLAS routines with mathematical-like syntax, and without undermining performance.
The BLAS implementation used at the backend could even run on graphics processing units (GPUs), e.g., by using the CUBLAS library for NVIDIA graphics cards.
Thus, through expression templates~\cite{Veldhuizen:1995}, the compiler can determine the precise function call at compilation time without adding extra running time overhead.
The methodology described herein is also available as an open-source header-only library for further reference~\cite{Aragon:2011}.
It is worth mentioning that both the Blitz++~\cite{Veldhuizen:1998} and the MTL~\cite{Gottschling:2007} \C++ libraries were built taking advantage of the C++ template facility. While the former was built entirely on the concept of expression templates, the latter uses generic programming techniques for achieving high performance.

This article is organized as follows:
Some of the \C++ features introduced by the new \CPP11 are summarized in Section~\ref{sec:C++11}.
Section~\ref{sec:algebra} shows the use of these paradigms in the implementation of an \code{Array} class template that can be used for the definition of tensors of any rank, i.e., a class template that can define vectors, matrices, fourth-order tensors, etc. 
By using operator overloading and expression templates, Section~\ref{sec:expr} sets forth the syntax that provides the user with an accessible mathematical notation.
It presents the implementation of the classes and operators that define the language that allows us to use the algebraic objects in mathematical expressions. Details on the usage of the implementation, together with a list of operations currently supported, are given in Section~\ref{sec:usage}. Finally, performance tests on both CPU and GPU are presented in Sections~\ref{sec:performance}.

\section{New \CPP11 language features} \label{sec:C++11}

Even though some of the components discussed in this article are known to any programmer who has been exposed to the \C++ programming language, other idioms require advanced knowledge, specially when dealing with metaprogramming, i.e., the generation of code at compilation time. Programmers usually think of a program as a set of instructions that do work at run time. However, some of the idioms used here will deal with compile time programming.
For the more advanced topics, the reader is referred to~\cite{Veldhuizen:1995, Alexandrescu:2001} and the references therein.

The main features from the \CPP11 set of requirements used in this work are illustrated in the following code snippets.
In this and the following sections, the \C++ code is given as concise as possible, and thus code that is not relevant (e.g., \keywordd{assert} statements) is removed to ease the presentation.
Prior to \CPP11, the \C++ programming language only supported a fixed number of template parameters for both classes and functions.
The introduction of \emph{variadic templates}~\cite{Gregor:2006, Gregor:2007} eliminates this restriction and brings unprecedented potential for template metaprogramming.
Thus, variadic templates are one of the main language components used in this paper:
%
%
\begin{lstlisting}[frame=none]
template <typename... Types> void foo(Types... args);

void bar() {}
template <typename T, typename... Types> void bar(T u, T v, Types... args) { 
  static_assert(sizeof...(args) $\%$ 2 == 0, "Odd number of parameters");
  cout<<u<<'-'<<v<<endl; bar(args...);
}

template <typename... Types> class ClassA;
template <typename T, typename... Types> class ClassB;
\end{lstlisting}
In the code snippet above, \code{Types} is a \emph{parameter pack identifier}, and the new syntax for variadic templates uses the ellipsis operator to indicate that functions \code{foo}, \code{bar}, and classes \code{ClassA}, \code{ClassB}, can take an arbitrary number of parameters. Notice the different placement of the ellipsis operator with respect to \code{Types} in the declarations above. Placing the operator to the left declares the identifier \code{Types} as a \emph{template parameter pack}, whereas placing it to the right denotes a \emph{parameter pack expansion}. In the declaration of the functions above, the pack expansion denotes a \emph{function parameter pack}. A pack expansion can be used in the body of a function, e.g., to produce a list of instantiated elements that match their corresponding type in the pack.
The function \code{bar} above uses variadic templates to print ordered pairs given an arbitrary even number of parameters. The definition of the function makes use of recursive calls with a pack expansion, and the recursion terminates with a call to the base case empty function. 
The \keywordd{sizeof}\code{...} operator, also introduced in the new standard, can be used to produce a compile time constant indicating the number of elements in a parameter pack. 

The \keywordd{static\_assert} keyword in the code snippet above represents a compile time assertion, and it has also been introduced in the new standard. 
By using this keyword, the compiler can determine errors at compilation time based on a boolean constant.
The concept of compile time assertions is not new, and it can be implemented with the previous standard~\cite{Alexandrescu:2001}.

\C++11 also introduces the ability to define \emph{alias templates}
\begin{lstlisting}[frame=none]
template <typename... Types>
using ClassBInt = B <int,Types...>;
\end{lstlisting}
so the resulting identifier \code{ClassBInt} represents a family of types that take an arbitrary number of template parameters but that have the first parameter of integral type \keywordd{int}.
All of these additions to the standard library are used in the next section to craft an efficient arbitrary-rank array class template.

\section{The \ttfont{Array} class template} \label{sec:algebra}

The objective of this section is to outline a single \code{Array} class template that would allow the creation of tensors of any rank.
In this way, an array for a specific rank can be determined with a simple alias template declaration
\begin{lstlisting}[frame=none]
template <int k, typename T> class Array; // array class template declaration

template <class T>
using vector_type = array::Array<1,T>; // family of vectors

template <class T>
using matrix_type = array::Array<2,T>; // family of matrices

template <class T>
using tensor_type = array::Array<4,T>; // family of 4th-order tensors
\end{lstlisting}

The preamble of the \code{Array} class template is shown in Listing~\ref{lst:array_template}.
\begin{lstlisting}[caption={ \code{Array} class template}, label=lst:array_template]
template <int k, typename T>
class Array : public Array_traits<k,T, Array<k,T> > {
  
public:
  
  typedef Array_traits<k,T, Array> traits_type;
  
  typedef T value_type;
  typedef T* pointer;
  typedef const T* const_pointer;
  typedef T& reference;
  typedef const T& const_reference;
      
private:
  
  size_t n_[k] = {0};
  pointer data_;
  bool wrapped_;
  
  // ...
\end{lstlisting}
Besides introducing some type definitions \emph{\`{a} la} \C++ standard library, the class template declares three member variables: the array dimensions are stored in \code{n\_}, the array elements in \code{data\_}, and the flag \code{wrapped\_} is used to indicate if the array owns its elements.

Because the interface of a tensor would depend on its actual rank, the \code{Array} class template is a subclass of \code{Array\_traits}, a class template that is displayed in Listing \ref{lst:array_traits}.
For example, a matrix would have functions to return the number of rows and columns, which do not make sense for a vector.  But the interface of the \code{Array\_traits} class template is limited to those functions that are specific to the particular rank, leaving most of the code to the common \code{Array} class.
\begin{lstlisting}[caption={ \code{Array\_traits} class template}, label=lst:array_traits]
template <int k, typename T, class array_type>
class Array_traits;

template <typename T, class array_type>
class Array_traits<1,T, array_type> {

public:
  T norm() const;
  
  // ... rest of the interface for a vector
};

template <typename T, class array_type>
class Array_traits<2,T, array_type> {

public:
  size_t rows() const;
  size_t columns() const;

  // ... rest of the interface for a matrix
};

// ... partial template specializations for higher-rank arrays
\end{lstlisting}
Thus, through partial template specialization, the developer can define the interface (and the state if needed) for tensors of a specific rank.
It is worth noting that the definition of the \code{Array\_traits} class template depends on the concrete \code{Array} type. This idiom, which is not a new addition to the language, 
is called the \emph{curiously recurring template pattern} (CRTP)~\cite{Coplien:1995}. Through this pattern, the partial template specializations of \code{Array\_traits} can access the variables of the concrete \code{Array} class through an appropriate type casting.

\subsection{Constructors}

The default and copy constructors, the assignment operator, and the destructor of the \code{Array} class template follow the traditional \C++ idioms.
Yet, the creation of vectors, matrices and other tensors from a single class template is possible due to the introduction of variadic templates~\cite{Gregor:2007}. The user can create algebraic objects by calling the same constructor, which takes a variable number of parameters depending on the rank of the object, as follows:
\begin{lstlisting}[frame=none]
vector_type<double> x(3);
matrix_type<float> A(3,3), B(3);
tensor_type<int> II(3,3,3,3), TT(3);
\end{lstlisting}

The constructor, shown in Listing~\ref{lst:variaic_constructor}, contains a compile time assertion to ensure that the number of parameters passed is at most equal to the rank of the array plus one. The extra parameter can be used to initialize all elements of the array with a default value, or to even use a lambda expression, or a functor, for the initialization.
\begin{lstlisting}[caption={Variadic template constructor}, label=lst:variaic_constructor]
  // inside Array class template
  template <typename... Args>
  Array(const Args&... args) : data_(nullptr), wrapped_() {    
    static_assert(sizeof...(Args) <= k+1, "*** ERROR *** Wrong number of arguments for array");
    init<0>(args...);
  }
\end{lstlisting}
The template function \code{init} allows us to analyze each of the parameters from the expansion individually. Depending on the type of the parameter, one of the functions in Listing~\ref{lst:init} is selected.
The \code{enable\_if} idiom was introduced by J\"arvi et al.~\cite{Jarvi:2003} to deal with ambiguous function calls in template overload resolution. That is, if the compile-time predicate written as the first parameter evaluates to false, the function template is not considered as a valid candidate for the function call due to the \emph{Substitution Error Is Not An Error (SFINAE)} situation. As a result, the compiler removes this function as a candidate instead of signaling a compilation error.
For example, the first function \code{init} in the listing would not be selected to be a valid function during overload resolution in any of the following three situations:: \emph{i)} the first parameter passed to the function is not of integral type; \emph{ii)} the first parameter is a pointer; and \emph{iii)} the recursive integer \code{d} is greater or equal than the tensor rank.
If none of these situations arise, this function then sets the corresponding dimension and calls itself recursively with the rest of the parameters.
\begin{lstlisting}[caption={Variadic constructor delegate function \code{init}}, label=lst:init]
  // init takes an integer parameter
  template <int d, typename U, typename... Args>
  typename std::enable_if<std::is_integral<U>::value and !std::is_pointer<U>::value and d < k, void>::type
  init(U i, Args&&... args) {
    n_[d] = i;
    init<d+1>(args...);
  }
      
  // init takes a value to initialize all elements
  template <int d>
  void init(value_type v = value_type()) {
    size_t s = init_dim();
    data_ = new value_type[s];
    std::fill_n(data_, s, v);
  }
  
  // init with a pointer to already existing data
  template <int d, typename P, typename... Args>
  typename std::enable_if<std::is_pointer<P>::value, void>::type
  init(P p, Args&&... args) {
    init_dim();
    wrapped_ = true;
    data_ = p;
  }
  
  // init takes a functor, lambda expression, etc.
  template <int d, class functor>
  typename std::enable_if<!std::is_integral<functor>::value and !std::is_pointer<functor>::value, void>::type
  init(functor fn) {
    size_t s = init_dim();
    data_ = new value_type[s];
    this->fill(fn);
  }
\end{lstlisting}

The \code{init} function that takes a \code{value\_type} as a parameter becomes active and finishes the recursion if all parameters are exhausted or if the last parameter is a valid value for the array elements.
In the former case, all elements in the array are initialized to zero, whereas in the latter they may be initialized to the passed parameter value.
The \code{init\_dim} initializes the dimensions of the array if necessary and returns the total number of elements. If not enough parameters are passed to the constructor, this function then copies the latest given size to the remainder dimensions. This is useful, e.g., to define a square matrix with a single parameter. 
The function then allocates the memory needed to store all elements in a one-dimensional array. The array elements are laid out in memory this way purposely to take advantage of the BLAS routines.

In case a pointer to memory is passed, the constructor does not allocate any memory and the \code{wrapped\_} flag is set to true. In this case the array does not own its memory, and thus the destructor of the class template takes this into account at the time to cleanup memory.

Finally, the final parameter can be a functor, or a lambda expression, that is used to initialize the elements of the array. For example, the following code can be used to create an identity matrix using this constructor:
\begin{lstlisting}[frame=none]
  matrix_type<double> A(5, [=](int i, int j) {return i == j ? 1 : 0;});
\end{lstlisting}
To accomplish this, the constructor calls the function \code{fill} of the base \code{Array\_traits} class.

\CPP11 introduces initializer lists so that objects can be constructed from an array of values of the same type.
An initializer list that is used to initialize an array would definitely depend on its rank. The following code uses initializer lists to create both a vector and a matrix.
\begin{lstlisting}[frame=none]
  vector_type<double> x = { 1., 2., 3., 4.};
  matrix_type<double> A = {{1., 2., 3.},{4., 5., 6.}};
\end{lstlisting}
The constructor in Listing~\ref{lst:init_list} is of parameter type \code{initializer\_type}, that is obtained at compilation time by referring to the type definitions within the helper class template \code{Initializer\_list}. In other words, during compilation the \code{initializer\_type} is obtained by a template recursion that is finished with \code{Initializer\_list<1,U>}. The same template recursion is used to set the elements of the array from the (possibly nested) initializer list.
\begin{lstlisting}[caption={Initializer list constructor}, label=lst:init_list]
  // helper local class defined inside Array
  template <int d, typename U>
  struct Initializer_list {

    typedef std::initializer_list<typename Initializer_list<d-1,U>::list_type > list_type;
    
    static void process(list_type l, Array& a, size_t s, size_t idx) {
      a.n_[k-d] = l.size(); // set dimension
      size_t j = 0;
      for (const auto& r : l)
        Initializer_list<d-1, U>::process(r, a, s*l.size(), idx + s*j++);
    }
  };
  
  // partial template specialization to finish recursion
  template <typename U>
  struct Initializer_list<1,U> {

    typedef std::initializer_list<U> list_type;

    static void process(list_type l, Array& a, size_t s, size_t idx) {
      a.n_[k-1] = l.size(); // set dimension
      if (!a.data_)  a.data_ = new value_type[s*l.size()];
      size_t j = 0;
      for (const auto& r : l)
        a.data_[idx + s*j++] = r;
    }
  };
  
  typedef typename Initializer_list<k,T>::list_type initializer_type;

  // initializer list constructor
  Array(initializer_type l) : wrapped_(), data_(nullptr)
  { Initializer_list<k, T>::process(l, *this, 1, 0); }
\end{lstlisting}

\subsection{Move semantics}

Prior to \CPP11, if a complex object were the result of a function, and if the body of the function were simple enough, then the compiler could perform the Return Value Optimization (RVO) to avoid the copy of the object being returned.
Yet, there were situations where the compiler could not accomplish such optimization, and thus \emph{move semantics}, a new feature of \CPP11, provide a very efficient way to deal with this problem.

Thus, in addition to the constructors defined in the previous section, the \code{Array} class template can benefit from move semantics so as to steal the \code{data\_} pointer that stores the array elements.
As such, the pointer of the temporary object is \emph{moved} to the new object.
The constructor and assignment operator that make this behavior possible are shown in Listing~\ref{lst:move}.
\begin{lstlisting}[caption={Move constructor and assignment operator}, label=lst:move]
  // move constructor
  Array(Array&& src) : data_(nullptr), wrapped_() {
    
    std::copy_n(src.n_, k, n_);
    data_ = src.data_;
    wrapped_ = src.wrapped_;
  
    src.data_ = nullptr;
    src.wrapped_ = false;
    std::fill_n(src.n_, k, 0);
  }
  
  // move assignment operator
  template <int k, typename T>
  Array& operator=(Array&& src) {  
    if (this != &src) {
      
      if (!wrapped_) delete data_;
    
      std::copy_n(src.n_, k, n_);
      wrapped_ = src.wrapped_;
      data_ = src.data_;
      
      src.data_ = nullptr;
      src.wrapped_ = false;
      std::fill_n(src.n_, k, 0);
    }
    return *this;
  }
\end{lstlisting}

\subsection{Element access}

Element access to the \code{Array} can be accomplished by overloading \keywordd{operator}\code{()} and \keywordd{operator}\code{[]}.
As before, variadic templates are used to provide element access through an overloaded \keywordd{operator}\code{()} that uses variadic template syntax, as illustrated in Listing~\ref{lst:variadic_operator()}.
\begin{lstlisting}[caption={Variadic template \code{operator()}}, label=lst:variadic_operator()]
  template <typename first_type, typename... Rest>
  struct Check_integral {
    typedef first_type pack_type;
    enum { tmp = std::is_integral<first_type>::value };
    enum { value = tmp && Check_integral<Rest...>::value };
    static_assert (value ,"*** ERROR *** Non-integral type parameter found.");
  };
  
  template <typename last_type>
  struct Check_integral<last_type> {
    typedef last_type pack_type;
    enum { value = std::is_integral<last_type>::value };
  };
  
public:    

  template <typename... Args>
  reference operator()(Args... params) {
    // check for correct number of parameters
    static_assert(sizeof...(Args) == k , "*** ERROR *** Number of parameters does not match array rank.");
    typedef typename Check_integral<Args...>::pack_type pack_type;
    pack_type indices[] = { params... };	// unpack parameters
    return data_[index(indices)];
  }
\end{lstlisting}
The function uses again a compile time assertion on the number of parameters passed. The helper class template \code{Check\_integral} defines the type of the parameter pack, and assures that all parameters are of integral type, yielding a compile-time error otherwise. The metaprogram works as before, by obtaining the type recursively and stopping with the partial template specialization for the last type. The latter is also the only helper class used in case of a vector, where a single parameter is passed to the function.
The function then calls a private member function to obtain the index in the one-dimensional array that contains the required element.
A similar function can be written for constant array objects, but it is not presented here for its definition is almost identical to that presented in Listing~\ref{lst:variadic_operator()}.

The traditional C-style means of accessing an array element through \keywordd{operator}\code{[]} can also be implemented for the \code{Array} class template. 
Because \keywordd{operator}\code{[]} is a unary operator, i.e., an operator that takes a single argument, the technique relays on  returning a proxy object from \keywordd{operator}\code{[]} when dealing with arrays for which \code{k>1}~\cite{Meyers:1996}. 
For example, if the array represents a matrix (i.e., \code{k=2}), the returned proxy object also implements \keywordd{operator}\code{[]} to return the element of the matrix. However, in our \code{Array} class template some metaprogramming is needed for the array has arbitrary rank. Listing~\ref{lst:proxy} shows the implementation of \keywordd{operator}\code{[]} and the generalized \code{Array\_proxy} class template.
\begin{lstlisting}[caption={Unary \code{operator[]} and proxy class template}, label=lst:proxy]
  // inside Array class
  typedef typename Array_proxy_traits<k,Array>::reference proxy_reference;
  typedef typename Array_proxy_traits<k,Array>::value_type proxy_value;
  
  proxy_reference operator[](size_t i)
  { return Array_proxy_traits<k,Array>::get_reference(*this,i); }
  proxy_value operator[](size_t i) const
  { return Array_proxy_traits<k,Array>::value(*this,i); }
  // ...
};

// outside Array class
template <int d, class Array>
struct Array_proxy_traits {
  
  typedef Array_proxy<d-1, Array> reference;
  typedef const Array_proxy<d-1, Array> value_type;
  
  static reference get_reference(Array& a, size_t i)
  { return reference(a,i); }
  static value_type value(Array& a, size_t i)
  { return value_type(a,i); }
};

template <class Array>
struct Array_proxy_traits<1,Array> {

  typedef typename Array::value_type primitive_type;
  typedef primitive_type& reference;
  typedef primitive_type const & value_type;
  
  static reference get_reference(Array& a, size_t i)
  { return a.data_[i]; }
  static value_type value(Array& a, size_t i)
  { return a.data_[i]; }
};

template <int d, class Array>
struct Array_proxy {
  
  typedef const Array_proxy<d-1, Array> value_type;
  typedef Array_proxy<d-1, Array> reference;
  
  // index set to first component of operator[], no further update needed
  explicit Array_proxy (const Array& a, size_t i)
  : a_(a), i_(i), s_(a.n_[0]) {}
  template <int c> Array_proxy (const Array_proxy<c, Array>& a, size_t i)
  : a_(a.a_), i_(a.i_ + a.s_ * i), s_(a.s_ * a.a_.n_[Array::rank() - c]) {}
  
  reference operator[](size_t i)
  { return reference(*this, i); }
  value_type operator[](size_t i) const
  { return value_type(*this, i); }
  
  const Array& a_;
  size_t i_, s_;
};

template <class Array>
struct Array_proxy<1, Array> {
  
  typedef typename Array::reference reference;
  typedef typename Array::value_type value_type;
  
  explicit Array_proxy (const Array& a, size_t i)
  : a_(a), i_(i), s_(a.n_[0]) {}
  template <int c> Array_proxy (const Array_proxy<c, Array>& a, size_t i)
  : a_(a.a_), i_(a.i_ + a.s_ * i), s_(a.s_ * a.a_.n_[Array::rank() - c]) {}
  
  reference operator[](size_t i)
  { return a_.data_[i_+i*s_]; }
  value_type operator[](size_t i) const
  { return a_.data_[i_+i*s_]; }
private:
  const Array& a_;
  size_t i_, s_;
};
\end{lstlisting}
There are two implementations of the unary \keywordd{operator}\code{[]}, one for reading from \keywordd{const}\code{ Array} objects and another for read/write access. 
The unary \keywordd{operator}\code{[]} inside the \code{Array} class returns an object of a type defined in the \code{Array\_proxy\_traits} class template. The latter defines this type as a proxy object of a lower rank, and the partial template specialization \code{Array\_proxy\_traits<1, Array>} finishes the compile-time recursion.

\subsection{Iterators}

Accessing an element through the operators defined above requires the computation of the index in the one-dimensional array that stores the elements of the array.
This computation can be completely avoided by providing iterators, as it is done for the containers of the \C++ standard library.
The proposed implementation defines an iterator class template \code{Array\_iterator}, declared as follows:
\begin{lstlisting}[frame=none]
template <typename T, typename P, int d=-1>
class Array_iterator;
\end{lstlisting}
where \code{T}, \code{P}, and \code{d} represent the value type, the pointer type, and the iterator dimension, respectively. The partial template specialization for \code{d=-1} advances the iterator by one in memory, e.g., as it is done for the the iterator of \code{std::vector}. This implementation is used to iterate linearly over all elements, regardless of its rank. The \code{Array} class template thus provides the commonly used functions \code{begin}, \code{end}, \code{rbegin}, and \code{rend} for both const and non-const objects.

Still, the general implementation for \code{d$\geq$0} shown in Listing~\ref{lst:iterator} provides a means to iterate over the different dimensions of the array. This implementation adds a member variable that saves the stride, i.e., the number of locations in memory that the pointer needs to be moved in memory to access the next element.
\begin{lstlisting}[caption=\code{Array\_iterator} class template, label=lst:iterator]
template <typename T, typename P, int d>
struct Array_iterator : public std::iterator<std::random_access_iterator_tag, T, ptrdiff_t, P> {
  
  typedef P pointer;
  
  Array_iterator(T* x, size_t str) :p_(x), str_(str) {}
  // ... other constructors

  Array_iterator& operator++()
  { p_ += str_; return *this; }
  
  Array_iterator operator++(int)
  { Array_iterator it(p_); p_ += str_; return it; }
  
  // ... other operators:
  // operator--, operator==, operator!=, operator*

private:
  pointer p_; 	// pointer
  size_t str_;	// stride
};
\end{lstlisting}

The \code{Array} class template then defines functions that contain as template parameter the dimension over which iteration is to be performed.
An alias template declaration within the class defines the dimensional iterator \code{diterator}. The functions that can make use of dimensional iterators are given in Listing~\ref{lst:iterator_functions}, hiding their counterparts for const arrays.
\begin{lstlisting}[caption={Functions for dimensional iteration}, label=lst:iterator_functions]
  template <int d>
  using diterator = Array_iterator<value_type, pointer, d>;
  
  template <int d> diterator<d> dbegin()
  { return diterator<d>(data_, stride(d)); }
  
  template <int d> diterator<d> dend()
  { size_t s = stride(d); return diterator<d>(data_ + stride(d+1), s); }
  
  template <int d, typename iterator>
  diterator<d> dbegin(iterator it) { return diterator<d>(&*it, stride(d)); }
  
  template <int d, typename iterator> diterator<d> dend(iterator it)
  { size_t s = stride(d); return diterator<d>(&*it + stride(d+1), s); }
\end{lstlisting}
As an example, the code below can be used to print each row of a matrix \code{A}.
\begin{lstlisting}[frame=none]
  int k = 0;
  typedef matrix_type<double>::diterator<0> row_iterator;
  typedef matrix_type<double>::diterator<1> col_iterator;
  for (row_iterator it1 = A.dbegin<0>(); it1 != A.dend<0>(); ++it1) {
    cout<<"row "<<k++<<':';
    for (col_iterator it2 = A.dbegin<1>(it1); it2 != A.dend<1>(it1); ++it2)
      cout<<' '<<*it2;
    cout<<'\n';
  }
\end{lstlisting}

\subsection{Standard output}

A similar metaprogram to those presented earlier can be written to output the array to a stream from the standard library.
Here, a class template \code{Print} implements a \code{print} static member function that calls the same function on a class of a lower dimension recursively at compilation time, until the recursion is finished with \code{Print<2>}. In the case of a vector object, the template specialization \code{Print<1>} handles the printing of the vector elements to the output stream.
%
%

The \code{Array} class template can have a much richer interface, and the metaprogramming techniques described above can be used to define it. If a function is coherent only for an array of specific rank, it can be defined in the corresponding partial template specialization of the \code{Array\_traits} class introduced earlier.

\section{Language expressions} \label{sec:expr}

This section layouts the syntax used to work with the \code{Array} class template presented in the previous section. The methodology follows that of expression templates, introduced by Veldhuizen~\cite{Veldhuizen:1995}.
Now that the \code{Array} class template is in place, it is desirable to write code as the following:
\begin{lstlisting}[frame=none]
matrix_type<double> A(m,n), B(n,m);
double alpha;
// initialize objects
// ...
matrix_type<double> C = alpha*A*B;
\end{lstlisting}
Listing~\ref{lst:expr_operators} presents the overloaded operators that are needed to accomplish the operation above.
An alias template is declared for the resulting expression of a scalar-array multiplication. In this way, the use of aliases reduces considerably the amount of code that the programmer has to write, and makes the code more readable and easier to maintain.
In the \code{SAm<d,T>} alias template declaration, \code{Expr} is a wrapper class template that contains the information about operators and operands without carrying out any computations.
As a result, the actual computation can be deferred to a later time when the result of the expression is needed. This technique, known as \emph{lazy evaluation}, is an important component for creating an efficient syntax for algebraic operations, as the expressions can be modified efficiently so that the least amount of work is done to obtain the required result.
As a simple example, the expression $\alpha \left( \beta \mathbf{A} \right)$, with scalars $\alpha, \beta$ and matrix $\mathbf{A}$, would be transformed into $\gamma\mathbf{A}, \gamma=\alpha\beta$ before evaluating the expression. Furthermore, the technique can be used to match patterns of complex mathematical expressions so that they can be handled using the higher-level BLAS routines.
\begin{lstlisting}[caption={Overloaded operators}, label=lst:expr_operators]
// scalar-array multiplication alias template
template <int d, class T>
using SAm = Expr<BinExprOp<ExprLiteral<T>, Array<d,T>, ApMul>>;
  
// operator*(scalar, array)
template <int d, typename S, typename T>
typename enable_if<is_arithmetic<S>::value, SAm<d,T>>::type
operator*(S a, const Array<d,T>& b) {
  typedef typename SAm<d,T>::expression_type ExprT;
  return SAm<d,T>(ExprT(ExprLiteral<T>(a),b));
}

// expression-(scalar-array multiplication) multiplication alias template
template <int d, class T, typename A>
using ESAmm = Expr<BinExprOp<Expr<A>, SAm<d,T>, ApMul>>;
    
// operator*(expr, array)
template<int d, typename T, class A>
ESAmm<d,T,A> operator*(const Expr <A>& a, const Array<d,T>& b) {    
    typedef typename ESAmm <d,T,A>::expression_type ExprT;
    return ESAmm<d,T,A>(ExprT(a, T(1)*b));
}
\end{lstlisting}
Note the use of the \code{enable\_if} structure presented earlier. If \code{S} is indeed an arithmetic type, the function template is a valid candidate for the overload resolution and the return type is defined to \code{SAm<d,T>}. The return type is constructed by the aid of an \code{expression\_type} type definition declared within the \code{Expr} class template, explained below in this section. 

The second alias template declares the resulting expression between the multiplication of an arbitrary expression and a \code{SAm<d,T>} expression.
 It is worth noting that one needs to implement only the operators for the base cases, i.e., those that deal with non-expression objects, as a simple operator taking two expressions automatically generates the required code. Listing~\ref{lst:expr_expr_mul} shows the definition of \keywordd{operator}\code{*} taking two expressions.
\begin{lstlisting}[caption={Overloaded multiplication operator for two expressions}, label=lst:expr_expr_mul]
// operator*(expr, expr)
template<class A, class B>
Expr<BinExprOp<Expr<A>, Expr <B>, ApMul>>
operator*(const Expr<A>& a, const Expr<B>& b) {
    
    typedef BinExprOp<Expr<A>, Expr<B>, ApMul> ExprT;
    return Expr<ExprT>(ExprT(a,b));
}
\end{lstlisting}

Now we turn our attention to the types that appear in Listings~\ref{lst:expr_operators} and \ref{lst:expr_expr_mul}. The \code{ExprLiteral} class template is used to wrap a constant or literal~\cite{Veldhuizen:1995}. The class template provides implicit conversion to the type stored. 
The \code{Expr} class template is shown in Listing~\ref{let:Expr}. It is used as a wrapper to a binary expression class template and it exists to minimize the number of operator overloads~\cite{Veldhuizen:1995}. The class declares type definitions of the wrapped class, of type \code{expression\_type}, and its interface redirects to the latter. The listing also shows the new optional return value syntax introduced by the standard, where the keyword \keywordd{auto} replaces the old return type and the actual return type is placed at the end of the function declaration. The type of the return type is obtained by using the \keywordd{decltype} operator, which was also introduced by the \CPP11 set of requirements to let the compiler infer the type of an expression.
\begin{lstlisting}[caption={\code{Expr} class template}, label=let:Expr]
// Expression wrapper class
template <class A>
class Expr {
   
    A a_;
        
public:
    
    typedef A expression_type;
    typedef typename A::result_type result_type;
    
    Expr() : a_() {}
    Expr(const A& x) : a_(x) {}
    
    auto left() const -> decltype(a_.left())  { return a_.left(); }
    auto right() const -> decltype(a_.right()) { return a_.right(); }

    operator result_type() { return a_(); }
    result_type operator()() const { return a_(); }
    
    friend inline std::ostream& operator<<(std::ostream& os,const Expr<A>& expr);
};
\end{lstlisting}

The \code{BinExprOp} class template, displayed in Listing~\ref{lst:BinExprOp}, encapsulates the actual binary expression. The class uses the \code{Expr\_traits} traits class to obtain the type of the left and right branches of the expression. This is due to the fact that some objects are better stored by reference so no copy is necessary when handling expressions. The \code{operator\_type} is the class that handles the work to be done on the operands and it will be explained in detail later in this section. The class template can also handle unary expressions, for the resulting expression can add an empty type as the right-branch. For example, the expression that declares the transposition of an array takes the form \code{Expr<BinExprOp<Array<d,T>, EmptyType, ApTr>>}, where \code{EmptyType} is an empty \keywordd{struct} whose sole purpose is to fill in where a template parameter is not needed~\cite{Alexandrescu:2001}.
\begin{lstlisting}[caption={\code{BinExprOp} class template}, label=lst:BinExprOp]
// binary expression class template
template <class A, class B, class Op>
class BinExprOp { 
    
    typename Expr_traits<A>::type a_;
    typename Expr_traits<B>::type b_;
    
public:
    
    typedef A left_type;
    typedef B right_type;
    typedef Op operator_type;
    typedef typename Return_type<left_type, right_type, operator_type>::result_type result_type;
    
    BinExprOp(const A& a, const B& b) : a_(a), b_(b) {}

    auto left() const -> decltype(a_) { return a_; }
    auto right() const -> decltype(b_) { return b_; }

    auto operator()() const -> decltype(Op::apply(a_, b_))
    { return operator_type::apply(a_, b_); } 
};
\end{lstlisting}

The only remaining component to discuss in detail is the operator applicative classes. The applicative multiplication class \code{ApMul} is given in Listing~\ref{lst:applicative}.
The class shows the functions needed to execute the code \code{matrix\_type$<$double$>$}\code{ C = alpha*A*B}. An alias template for the multiplication between a scalar and a matrix is declared using the scalar-array alias template of Listing~\ref{lst:expr_operators}.
The static function shown first in the listing is used to carry out the multiplication between two matrices, each one with its respective scalar. In other words, this function can be used to obtain $\mathbf{C} = \left( \alpha \mathbf{A} \right) \left( \beta \mathbf{B} \right)$. In this implementation, the function call obtains the result of the multiplication by calling the respective BLAS function, but developers are allowed to call another function or to provide their own implementation.
However, since there is an optimized library that implements the BLAS interface in almost every computer architecture, it is strongly recommended to make calls to the BLAS routines.
A catch-all template function is also needed, so that the multiplication between unknown expressions calls \keywordd{operator}\code{()} recursively until a base case, as the one shown earlier for the multiplication between matrices, finishes the recursion.
As the syntax of the expressions evolves, similar static functions can be added to the applicative operator classes to provide further functionality.
\begin{lstlisting}[caption={\code{ApMul} multiplication applicative class}, label=lst:applicative]
// scalar - matrix multiplication alias template
template <typename T>
using SMm = SAm<2,T>;

// Multiplication applicative class
class ApMul {
public:
    
    // ... other applicative functions
    
    // SVm - SVm multiplication
    template <typename T>
    static matrix_type<T> apply(const SMm <T>& x, const SMm <T>& y) {
    
      // get matrix references
      const matrix_type<T>& a = x.right();
      const matrix_type<T>& b = y.right();
      
      matrix_type<T> r(a.rows(), b.columns());
      cblas_gemm(CblasNoTrans, CblasNoTrans, r.rows(), r.columns(),
               a.columns(), x.left()*y.left(),
               a.data_, a.rows(), b.data_, b.rows(), 1.0, r.data_, r.rows());
      return r;
    }
    
    // ... other applicative functions

    // expr - expr multiplication
    template<class A, class B>
    static typename Return_type<Expr<A>, Expr<B>, ApMul>::result_type
    apply(const Expr<A>& a, const Expr<B>& b)
    { return a()*b(); }
};
\end{lstlisting}

Listing~\ref{lst:Return_type} shows the metaprogram used to obtain the \code{Return\_type} of an expression, which has already been used in Listings~\ref{lst:BinExprOp} and \ref{lst:applicative}. The first four definitions of the \code{Return\_type} class template are general and deal with the operations of expressions and binary expression objects.
When dealing with a complicated expression that encapsulates many more complex expressions, the metaprogram uses these general classes to obtain the return type of the wrapped objects recursively at compile time. 
The recursion ends with the partial template specializations that deal with operations between base objects, e.g., an array or a literal.
\begin{lstlisting}[caption={\code{Return\_type} metaprogram}, label=lst:Return_type]
template <typename... Params>
struct Return_type;

// return type for an arbitrary binary expression with arbitrary operation
template <typename A, typename B, class Op>
struct Return_type < Expr < BinExprOp <A, B, Op > > > {
    typedef typename Return_type<A>::result_type left_result;
    typedef typename Return_type<B>::result_type right_result;
    typedef typename Return_type<left_result, right_result, Op >::result_type result_type;
};

// return type for the operation of an arbitrary object with an arbitrary expression
template <typename A, typename B, class Op>
struct Return_type < A, Expr < B >, Op > {
    typedef typename Return_type<typename B::left_type,
    typename B::right_type, typename B::operator_type>::result_type right_result;
    typedef typename Return_type<A, right_result, Op >::result_type result_type;
};

// return type for the operation of an arbitrary expression with an arbitrary object 
template <typename A, typename B, class Op>
struct Return_type < Expr<A>, B, Op > {
    typedef typename Return_type<typename A::left_type,
    typename A::right_type, typename A::operator_type>::result_type left_result;
    typedef typename Return_type<left_result, B, Op >::result_type result_type;
};

// return type for the operation between two arbitrary expressions 
template <typename A, typename B, class Op>
struct Return_type < Expr<A>, Expr<B>, Op > {
    typedef typename Return_type<typename A::left_type,
    typename A::right_type, typename A::operator_type>::result_type left_result;
    typedef typename Return_type<typename B::left_type,
    typename B::right_type, typename B::operator_type>::result_type right_result;
    typedef typename Return_type<left_result, right_result, Op >::result_type result_type;
};

// partial template specializations for scalar - matrix operation
template <int d, typename T, class Op>
struct Return_type<ExprLiteral<T>, Array<d, T>, Op> {
    typedef Array<d, T> result_type;
};

// transposed vector - vector multiplication
template <typename T>
struct Return_type<BinExprOp<Array<1,T>, EmptyType, ApTr>, Array<1,T>, ApMul> {
    typedef T result_type;
};

// vector - transposed vector multiplication
template <typename T>
struct Return_type<Array<1,T>, BinExprOp<Array<1,T>, EmptyType, ApTr>, ApMul> {
    typedef Array<2,T> result_type;
};
// ... further partial template specializations
\end{lstlisting}
For example, the return type of an operation between a scalar and a matrix defines the matrix as the return type. This does not mean that we can add a scalar to a matrix, as the applicative class has to define the supported operations. The following partial template specialization in the listing defines the return type for the multiplication between a transposed vector and a vector, i.e., for a scalar product. The final partial template specialization defines the return type for the dyadic product between two vectors. Many other specializations can be implemented depending on the needs of the developer.

There is one more point that needs to be discussed, as the example above produces a newly created matrix that holds the result of the multiplication.
Nevertheless, it is often required the modification of an existing object with the result of an expression, instead of the creation of a new one.
In \C++ this is traditionally accomplished on built-in types by using the compound assignment operators, i.e., operators \code{+=, -=, *=, /=}. Listing~\ref{lst:compound_assignment} shows the code needed to modify a matrix with the result of an expression that involves scalar and matrix multiplications, i.e., to execute the code \code{C += (alpha*A)*(beta*B)}.
It is worth noting that there is no running time performance loss as all the expression template facility is bound to compilation time.
\begin{lstlisting}[caption={Addition compound assignment operator}, label=lst:compound_assignment]
// operator+=(any, any)
template <class A, class B>
typename
enable_if<!is_arithmetic<A>::value && !is_arithmetic<B>::value, A& >::type
operator+=(A& a, const B& b) {
    typedef RefBinExprOp<A, B, ApAdd> ExprT;
    return Expr<ExprT>(ExprT(a,b))();
}

// (scalar - matrix multiplication) - (scalar - matrix multiplication) multiplication alias template1
template <typename T>
using SMmSMmm = Expr<BinExprOp<SMm<T>, SMm<T>, ApMul>>;

// Addition applicative class
class ApAdd {
public:
    
    // ... other applicative functions
    
    // array - (scalar*array - scalar*array multiplication) addition
    template <typename T>
    static matrix_type<T>& apply(matrix_type<T>& c, const SMmSMmm <T>& y) {
        
        // get matrix references
        const matrix_type<T>& a = y.left().right();
        const matrix_type<T>& b = y.right().right();
                
        cblas_gemm(CblasNoTrans, CblasNoTrans, a.rows(), b.columns(),
               a.columns(), y.left().left()*y.right().left(),
               a.data_, a.rows(), b.data_, b.rows(), 1.0, c.data_, c.rows());
        return c;
    }
    // other applicative functions
};
\end{lstlisting}

\section{Usage} \label{sec:usage}

The following listing shows an example of the mathematical syntax outlined in the previous section.
Clearly, the code is easier to read than writing calls to the BLAS routines. 
\begin{lstlisting}[caption={Example usage}, label=lst:usage]
#include <iostream>
#include "expr.hpp"

using namespace array;

int main() {
  
  int m = 4, n = 8;
  double alpha = 0.5, beta = 4.;
  vector_type<double> x(n,1.), y(n,1.);
  matrix_type<double> A(m,n,1.), B(m,n,1.);
  
  // initialize objects
  // ...
   double s = transpose(x)*y; // scalar product
  matrix_type<double> C = x*transpose(y); // outer product
  x += C*y; // matrix-vector multiplication
  C += alpha*transpose(A)*beta*B; // matrix-matrix multiplication
  
  std::cout<<"s: "<<s<<std::endl;
  std::cout<<"x: "<<x<<std::endl;
  std::cout<<"C: "<<C<<std::endl;
    
  return 0;
}  
\end{lstlisting}

The following table summarizes the operations currently supported by the framework.

\begin{table}
\centering
\begin{tabular}{>{}m{4cm} >{}p{10cm} >{}p{1.5cm}}
\rowcolor{Gray}
\textcolor{white}{math syntax} & \textcolor{white}{code} & \textcolor{white}{function} \tabularnewline
\hline 
\hline 
\rowcolor{lightergray}
Level 1 BLAS & & \tabularnewline
$y = x$ & \code {y = x;} & \code{\textcolor{red}{x}}\code{copy}  \tabularnewline

$y = \alpha x + y$ & \code {y += alpha * x;} & \code{\textcolor{red}{x}}\code{axpy}  \tabularnewline

$\alpha = x^\intercal y$ & \code {alpha = transpose(x) * y;} & \code{\textcolor{red}{x}}\code{dot}  \tabularnewline

$\alpha = \left\Vert x \right\Vert $ & \code {alpha = x.norm();} & \code{\textcolor{red}{x}}\code{nrm2}  \tabularnewline
\hline 
\rowcolor{lightergray}
Level 2 BLAS & & \tabularnewline
$y = \alpha A x + \beta y$ & \code {y += alpha * A * x + beta * y;} & \code{\textcolor{red}{x}}\code{gemv}  \tabularnewline

$y = \alpha A^\intercal x + \beta y$ & \code {y += alpha * transpose(A) * x + beta * y;} & \code{\textcolor{red}{x}}\code{gemv}  \tabularnewline

$A = \alpha x y^\intercal$ & \code {A = alpha * x * transpose(y);} & \code{\textcolor{red}{x}}\code{ger}  \tabularnewline

\hline 
\rowcolor{lightergray}
Level 3 BLAS & & \tabularnewline
$C = \alpha A B + \beta C$ & \code {C += alpha * A * B + beta * C;} & \code{\textcolor{red}{x}}\code{gemm}  \tabularnewline
$C = \alpha A^\intercal B + \beta C$ & \code {C += alpha * transpose(A) * B + beta * C;} & \code{\textcolor{red}{x}}\code{gemm}  \tabularnewline
$C = \alpha A B^\intercal + \beta C$ & \code {C += alpha * A * transpose(B) + beta * C;} & \code{\textcolor{red}{x}}\code{gemm}  \tabularnewline
$C = \alpha A^\intercal B^\intercal + \beta C$ & \code {C += alpha * transpose(A) * transpose(B) + beta * C;} & \code{\textcolor{red}{x}}\code{gemm}  \tabularnewline

\hline 
\end{tabular}
\caption{Operations currently supported by the framework. In the right column, the \code{\textcolor{red}{x}} can be replaced by either \code{s} or \code{d} for \keywordd{float} and \keywordd{double} data types, respectively.}
\end{table}

\section{Performance} \label{sec:performance}

\begin{figure}
 \centering 
        \subfloat[]{\label{fig:blas_ddot}{\includegraphics[scale=0.4]{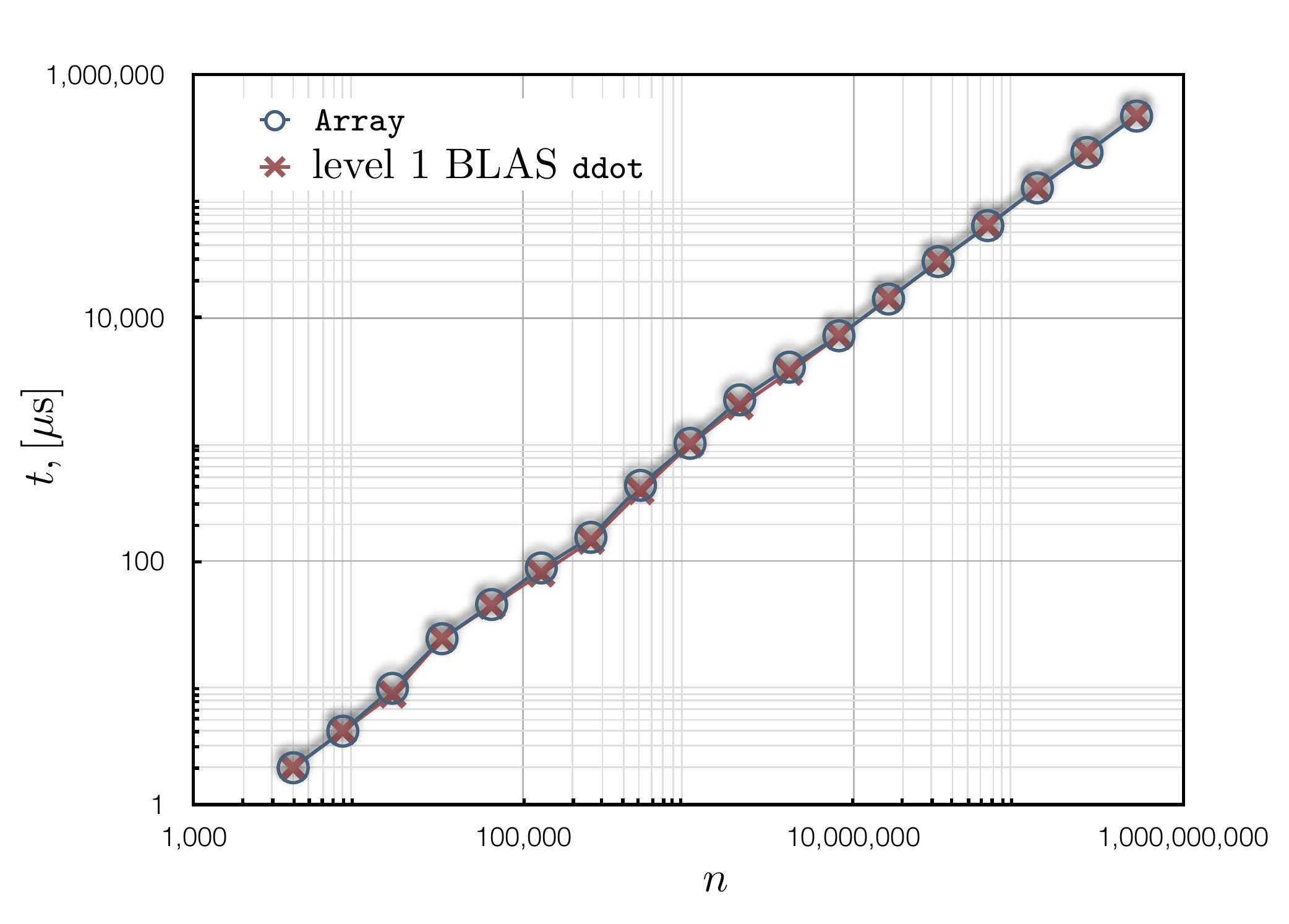}}} \\
        \subfloat[]{\label{fig:blas_dgemv}{\includegraphics[scale=0.4]{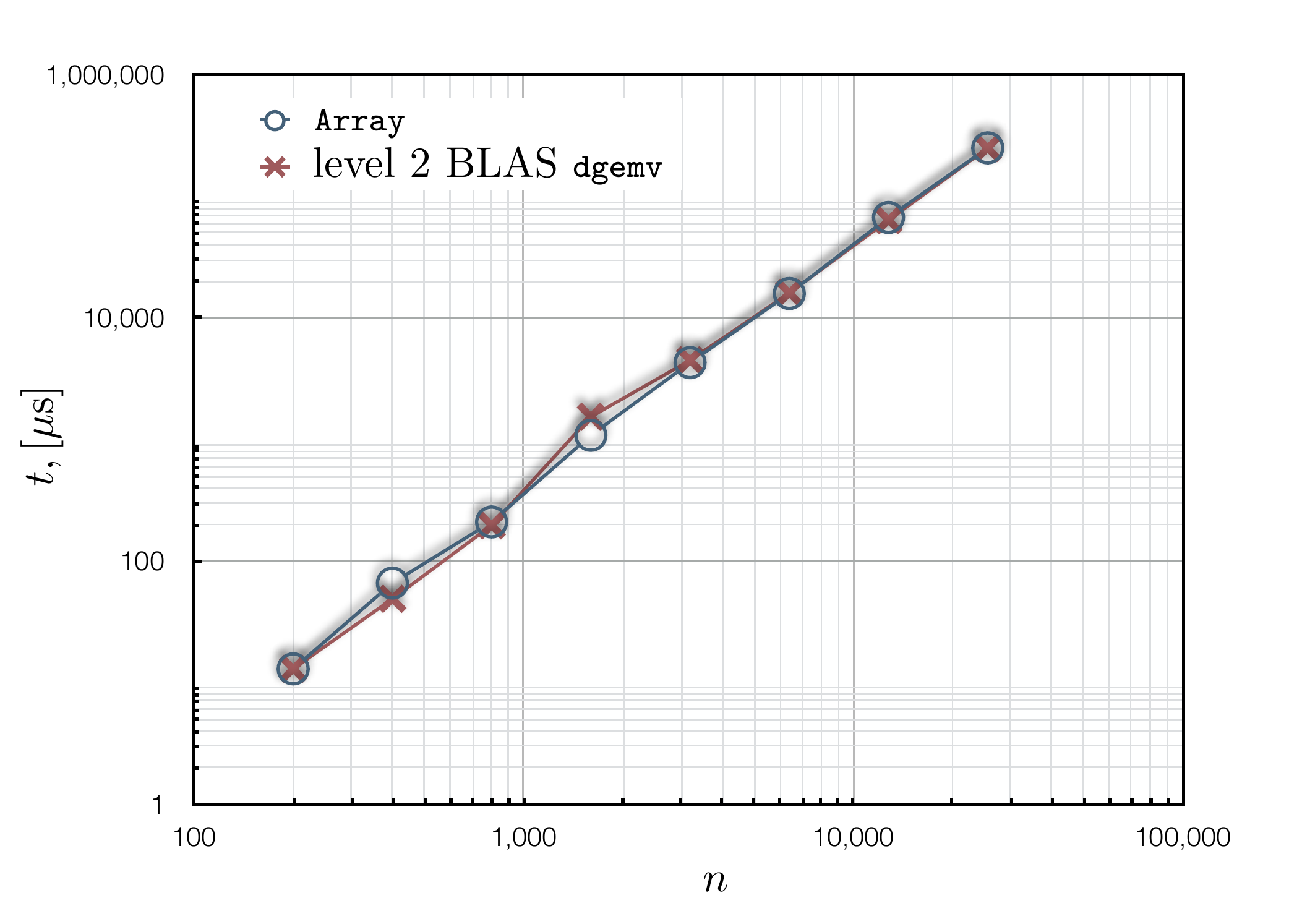}}} \\
        \subfloat[]{\label{fig:blas_dgemm}{\includegraphics[scale=0.4]{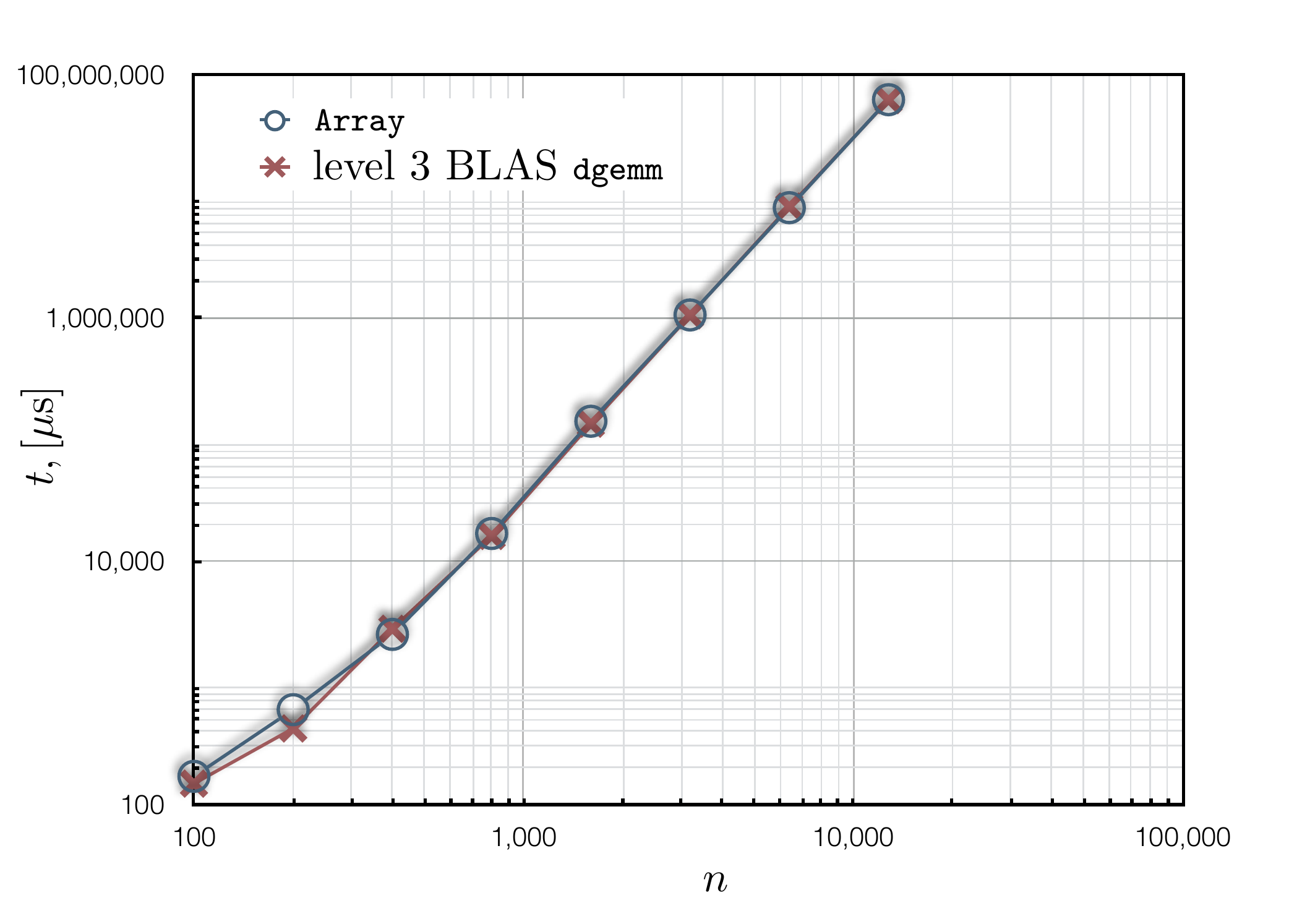}}}
  \caption{\label{fig:blas}Execution time $t$ (in $\mu$s) as a function of matrix size $n$, average of 10 operations of type: (a)  scalar product (\code{ddot}); (b) matrix-vector multiplication (\code{dgemv}); and (c) matrix-matrix multiplication (\code{dgemm}). The two curves compare the approach described using expression templates and the \code{Array} class described in the previous sections, with direct calls to the corresponding BLAS functions.}
\end{figure}

Performance tests were conducted using the \code{Array} class template described in Section~\ref{sec:algebra} with the mathematical syntax outlined in Section~\ref{sec:expr}. Comparisons were made with direct calls to the corresponding BLAS functions.
The compiler used was Clang version 5.0, and the programs were executed on an Apple MacBook Pro with a 2.6 GHz Intel Core i7 processor.
The programs linked to the BLAS library within the \code{Accelerate.framework} for Mac OS X.
For each matrix size, the reported time is the result of the average of ten different computations.
Figure~\ref{fig:blas_ddot} compares the code with a direct call to the level 1 BLAS function \code{ddot}, i.e., for the scalar product between two vectors. Matrix-vector multiplication (level 2 BLAS function \code{dgemv}) results are given in Figure~\ref{fig:blas_dgemv}. Finally, Figure~\ref{fig:blas_dgemm} shows the results for matrix-matrix multiplication, i.e., the level 3 BLAS function in Listing~\ref{lst:blas_dgemm}.
As the figures show, no performance is lost by using the presented framework.

\begin{figure}
 \centering 
        \subfloat[]{\label{fig:cublas_dgemm_a}{\includegraphics[scale=0.4]{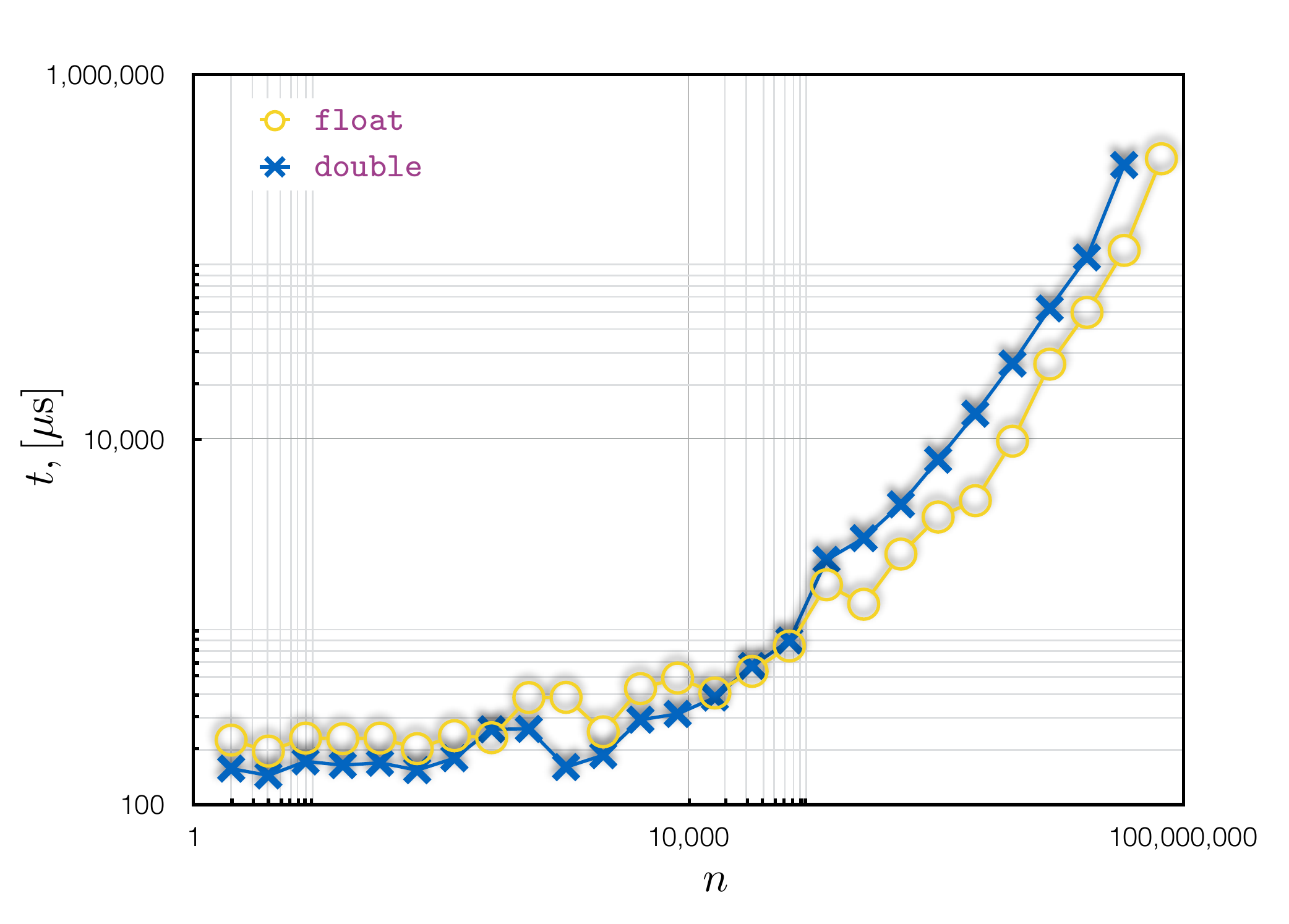}}} \\
        \subfloat[]{\label{fig:cublas_dgemm_b}{\includegraphics[scale=0.4]{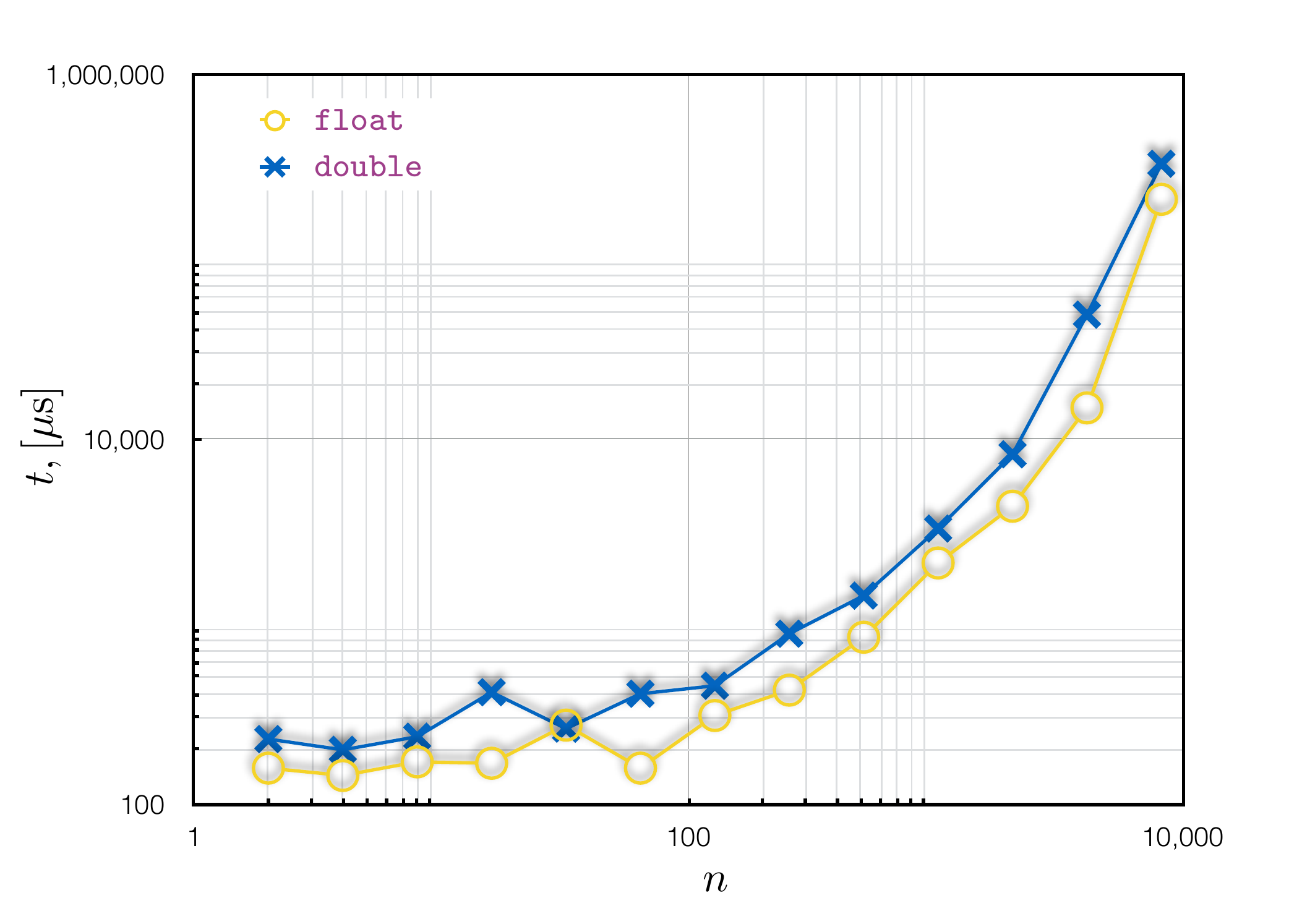}}} \\
        \subfloat[]{\label{fig:cublas_dgemm_c}{\includegraphics[scale=0.4]{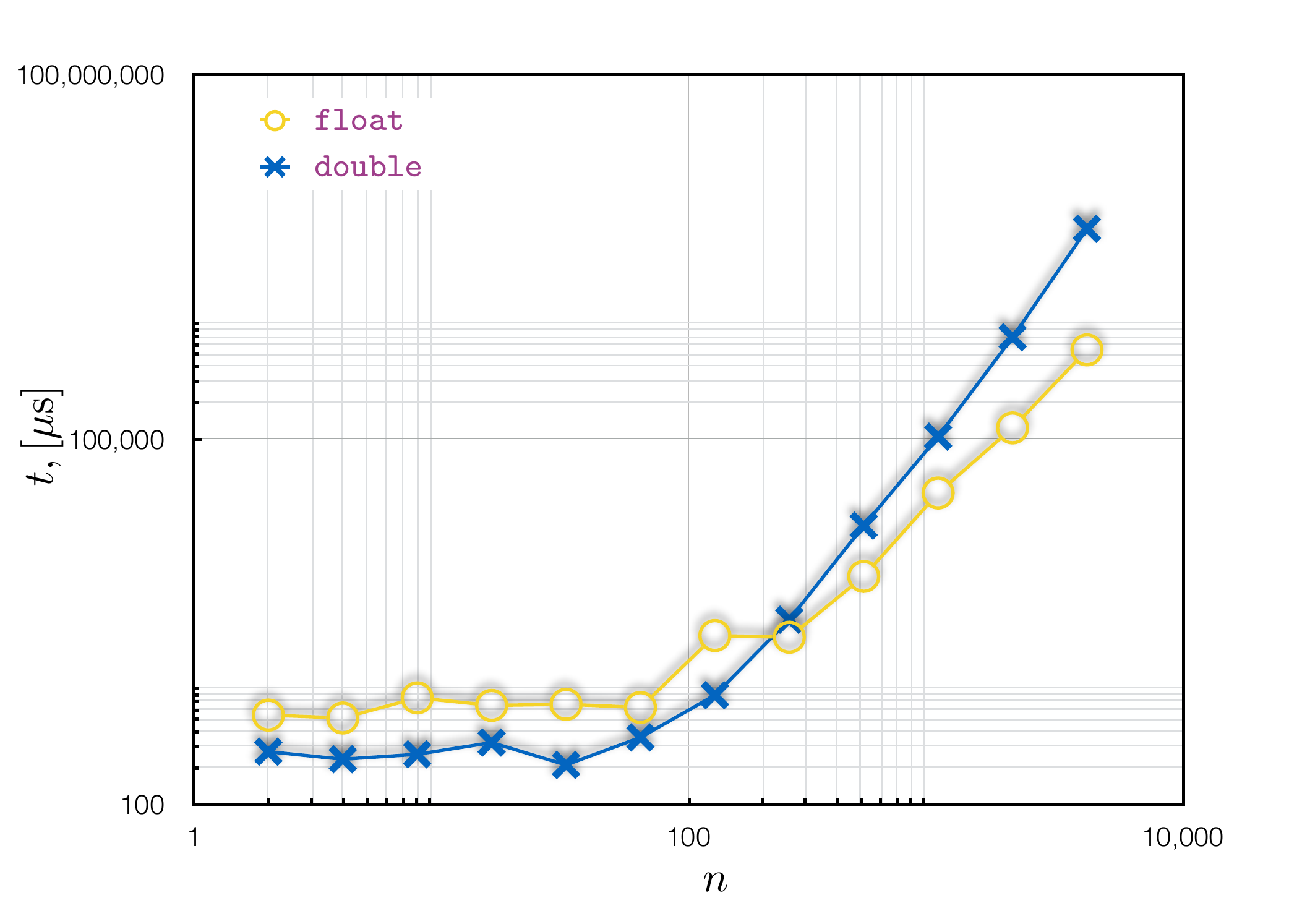}}}
  \caption{\label{fig:cublas_dgemm}Execution time $t$ (in $\mu$s) as a function of matrix size $n$, average of 10 operations of type: (a)  scalar product (\code{dot}); (b) matrix-vector multiplication (\code{gemv}); and (c) matrix-matrix multiplication (\code{gemm}). The results were obtained for both \keywordd{float} and \keywordd{double} data types using the CUBLAS library on a GeForce GT 650M.}
\end{figure}

Figure~\ref{fig:cublas_dgemm} shows the results of carrying out the same operations, but this time using the NVIDIA CUBLAS library at the backend. Results are reported for both \keywordd{float} and \keywordd{double} data types, obtained with an NVIDIA GeForce GT 650M graphics card. As apparent from the figure, it is not worth running these operations on GPUs for small vector and matrix sizes due to the time spent in transferring the data to and from the GPU.

\section{Conclusion} \label{sec:conclusion}

This article introduces many of the new features provided by the new \C++ standard in implementing efficient arbitrary-rank arrays, the building blocks of any scientific computing system.
It was shown how the new idioms were used to craft an efficient \code{Array} class template, that can generate tensors of any rank by using the same code.
By using operator overloading and expression templates, a straightforward mathematical syntax is provided, and the computation is deferred to a time where the result of the expression is actually needed. Thus, an extra high-level layer is given to the \C++ programming language when dealing with algebraic objects. As a result, the burden to the end user is drastically reduced, as the mathematical syntax can conceal the details of how operations are implemented behind the scenes. Tests conducted show that there is no loss of performance when using the proposed framework. Furthermore, it was shown that using the NVIDIA CUBLAS library at the backend is not worth it when the sizes of the involved matrices and vectors are small.





\bibliographystyle{unsrt}
\bibliography{../bibliography}

\end{document}